\documentclass[aps,pre,floatfix,twocolumn,nofootinbib,showpacs]{revtex4} 
\usepackage{graphicx} \usepackage{amsmath} \usepackage{amssymb}
\newcommand{\BEQ}{\begin{equation}}
\newcommand{\EEQ}{\end{equation}}
\newcommand{\BEA}{\begin{eqnarray}}
\newcommand{\EEA}{\end{eqnarray}}

\renewcommand{\d}{{\rm d}}

\renewcommand{\S}{{\rm S}}
\newcommand{\F}{{\rm F}}

\newcommand{\G}{\Gamma}

\newcommand{\dtau}{\frac{\d}{\d \tau}}

\newcommand{\CH}{{\cal H}}

\newcommand{\CP}{{\cal P}}

\newcommand{\C}{{\rm C}}
\newcommand{\E}{{\rm E}}

\begin{document}
\draft
\title
{Relating the thermodynamic arrow of time to the causal arrow.}
\date{\today}
\author{Armen E. Allahverdyan$^{1)}$ and Dominik Janzing$^{2)}$}
\affiliation{
$^{1)}$Yerevan Physics Institute, Alikhanian Brothers Street 2, Yerevan 375036, Armenia}
\affiliation{$^{2)}$School of Electrical Engineering and Computer Science,
University of Central Florida, Orlando, FL 32816-2362 }

\begin{abstract} Consider a Hamiltonian system that consists of a slow
subsystem S and a fast subsystem F. The autonomous dynamics of S is driven by an
effective Hamiltonian, but its thermodynamics is unexpected.  We show
that a well-defined thermodynamic arrow of time (second law) emerges for
S whenever there is a well-defined causal arrow from S to F and the
back-action is negligible.  This is because the back-action of F on S is
described by a non-globally Hamiltonian Born-Oppenheimer term that
violates the Liouville theorem, and makes the second law inapplicable to S.  
If S and F are mixing, under the causal
arrow condition they are described by microcanonic distributions $P(\S)$
and $P(\S |\F)$. Their structure supports a causal inference principle
proposed recently in machine learning.

\end{abstract}



\maketitle

\section{Introduction.}

In this paper we establish a relation between the causal arrow|i.e.,
emergence of a unidirectional interaction between two interacting
systems|and the thermodynamic arrow of time. Studying causation in the
context of various physical arrows of time is not a new subject
\cite{Reichenbach,Penrose,gold}. One of the motivations for these
studies is the analogy between the temporal asymmetry implied by the
thermodynamic arrow and the asymmetry between the cause and effect:
causes influence their effect, but not vice versa, and causes can only
happen {\it before} their effects \cite{Reichenbach,Penrose,gold}. 

Causal explanations in everyday-life often construct causal
structures among phenomena that are not well-localized in time (e.g., if
one studies relations between crime and poverty in social sciences). Even
for this kind of phenomena we observe sometimes well-defined causal
connections where one phenomenon is the cause and another one the
effect.  For understanding the link between thermodynamics and
causality within a statistical physics setting, it is helpful to study
the conditions under which we can consider one of two interacting
systems as the cause and the other the as effect. The question is then
to what extent the unidirectionality of the influence is related to the
thermodynamics of the two systems. 

The presented results provide some answers to the above general
question.  For describing those answers we proceed with separate
introductions on the thermodynamic arrow and the causal arrow.  This
section then closes with qualitative discussions of our main results. 

\subsection{The thermodynamic arrow of time.}

Thermodynamic arrow of time refers to formulations of the second law.
The understanding of this law from the first principles of quantum or
classical dynamics is achieved within statistical physics (in contrast
to thermodynamics, where the second law is postulated).  In this
statistical physics context we list the following basic formulations of
the second law \cite{Balian,Lindblad,Lenard}:

\begin{itemize}

\item Entropy formulation: coarse-grained entropy does not decrease in
time for a closed system that starts to evolve from a certain 
non-equilibrium state \cite{Balian,Jaynes,Lindblad,Zeh}. 

\item Thomson's formulation: for a system that starts to evolve from an
equilibrium state, no work extraction is possible by means of a cyclic
process driven by an external source of work \cite{Lindblad,Lenard}. 

\end{itemize}

These statements entail an arrow of time, since they refer to the
difference between final and initial values of the entropy and energy,
\footnote{Since any interaction with an external source of work can be
seen as a thermally isolated process, work is a difference between
average energies; see section \ref{trali} for details.} respectively.
Each formulation has two different aspects: special {\it initial}
conditions (non-equilibrium states for the entropy formulation,
equilibrium states for Thomson's formulation) and specific dynamic
features of the system (closed dynamics, cyclic processes).  Both these
aspect were studied from the first principles
\cite{Balian,Jaynes,Zeh,Lindblad,Lenard,Dom_Beth,mwp,Penrose,Reichenbach,Mahler,cosmology}
\footnote{The fact that we impose initial, and not final, conditions cannot be
derived from the first principles. Instead, it should be taken as a fact
that experiments are described by their {\it initial} conditions rather
than being described by the {\it final} conditions.}. 

There are more formulations of the second law, such as the minimal work
principle \cite{Balian,mwp,campisi} or the Clausius formulation
\cite{Balian,Dom_Beth,cl}.  Formulations of the second law are not always
equivalent \cite{cl,mwp}. The Thomson and entropy formulations do not require
anything more than a Hamiltonian dynamics that satisfies the Liouville
equation \cite{Balian,Lindblad,Lenard}, while the minimum work principle and the Clausius
formulation
do have additional requirements: ergodic observable of work for the
minimum work principle \cite{mwp} and weak (or conserved) interaction
Hamiltonian for the Clausius formulation \cite{Balian,Dom_Beth,cl}.  

We shall thus focus on the Thomson and entropic formulations. Here the preference
should be given to Thomson's formulation, since there is no universally
accepted definition of physical entropy for non-equilibrium states.  In contrast,
there is such a definition for work \cite{Balian,Lindblad}.  The
formulation and derivation of the entropy and Thomson's formulation will
be recalled below in section \ref{2law2}. 

\subsection{The causal arrow.}
\label{causal_intro}

Causal arrow refers to a dynamical situation when one variable $\C$
(cause) influences on another $\E$ (effect), but does not get
back-reaction \footnote{By the causal arrow we thus do not mean the
macroscopic causality that is when the past macro-state determines the
future one.}.  In this context we shall recall two operational
definitions of the causal arrow: {\it i)} Cutting off the interaction
between $\C$ and $\E$ does not alter the dynamics of $\C$, while it
influences the dynamics of $\E$. {\it ii)} Perturbing the
dynamics of $\C$|e.g., by means of external fields, or by changing the
initial conditions of $\C$| will influence the dynamics of $\E$, while
perturbing the dynamics of $\E$ will not influence on $\C$.

In studying causal relations (e.g. in economy, medicine, social
sciences), scientific reasoning often depends on statistical data that
has been obtained from mere observations. This is because interventions
that would {\it prove} causal relations are often impossible. One then
tries to draw plausible causal conclusions merely from stochastic
dependences in the joint distribution function $P(\C,\E)$ of the
variables \cite{Pearl}.  In spite of their obvious importance|as
sometimes our very survival depends on the proper identification of the
cause versus effect|such conclusions cannot be always correct, they are
merely plausible in the sense that they lead to correct predictions more
frequently than they fail \footnote{The fact that stochastic dependences
cannot serve as the basis for drawing unique causal conclusions was
stressed by Hume \cite{hoff}.}. 

Several criteria are known for this type of causal reasoning, if the
number of variables involved in a network of possible causal relations
is three or more \cite{Pearl}. The case of two variables is the most
difficult one, since there are no widely accepted causal reasoning
criteria for this situation.  For this case it was recently proposed
that one can plausibly identify $\C$ as the cause, if the probability
distributions $P(\E|\C)$ and $P(\C)$ are in a certain sense simpler than
$P(\C|\E)$ and $P(\E)$, respectively \cite{Jan}. 
Note that the ideas in \cite{Kano} can be interpreted in the same spirit.

\subsection{Purposes and results of the present work.}

{\bf 1.} We shall follow in detail how the causal arrow and the
thermodynamic arrow of time emerge in a closed, classical Hamiltonian
system that consists of two subsystems S and F.  For the sake of
studying causal arrow it is natural to assume that S is slow, while F is
fast. 

In a more general perspective, the idea of slow versus fast variables
has been developed in several different contexts, e.g., the
slaving principle proposed by Haken as a cornerstone for synergetics,
self-organization, and hierarchical dynamics \cite{Haken}. Indeed, many
(almost all?) models studied in mechanics, (non)equilibrium statistical
physics, chemical kinetics, mathematical ecology, {\it etc}, are not
fundamental, but rather describe effective behavior of slow degrees of
freedom. 

{\bf 2.}
The absence of the causal arrow in the above closed system is
quantified by the back-reaction of $\F$ on $\S$.  Under some natural
conditions outlined below, this back-reaction amounts to an additional
(Born-Oppenheimer\footnote{The names come from the early days
of atomic physics, when M. Born and R. Oppenheimer calculated in the quantum
mechanical setting the force exerted by fast electrons on slow nuclei.}) 
term in the Hamiltonian of $\S$.  The dynamics of
$\S$ is then autonomous and energy-conserving.  However, the
Born-Oppenheimer term has the following peculiar feature: it depends
explicitly on the initial value of the 
coordinates of $\S$ that participate in the 
interaction with $\F$.  This is a consequence of memory generated during
the tracing out of the fast variables.  Thus there is no single, global
Hamiltonian for $\S$.  We shall show that due to this fact the basic
formulations of the second law do not apply to $\S$, even if we assume
the existence of proper initial conditions. The reasons for this
inapplicability are discussed in detail in section \ref{trali}.
The main reason is that the Liouville theorem (conservation of the
phase-space volume) does not apply to $\S$. 
Thus, the usual formulation of the thermodynamic arrow of time
does not apply to $\S$ \footnote{This does not mean that
there cannot be other|apart from the thermodynamic arrow in the sense 
explained in the introduction|temporal
asymmetries in the dynamics of $\S$.}. 

{\bf 3.} If the Born-Oppenheimer term can be neglected for the dynamics
of $\S$, the applicability of the second law for $\S$ is recovered. This
neglection indicates on the existences of the causal arrow in the considered system: $\S$
appears to be the cause for $\F$. Thus the local
thermodynamic arrow for $\S$ emerges {\it due} to the causal arrow. 

Note that the second law applies to the fast subsystem
F, which has a driven, globally Hamiltonian dynamics. Such a dynamics
serves as a basis for deriving the second law from the first principles
\cite{Balian,Lindblad,Lenard,Dom_Beth,mwp,campisi}. 

{\bf 4.} Another important consequence of the Born-Oppenheimer term is
that it makes $\S$ strongly non-ergodic, even if the bare Hamiltonian of
$\S$ is assumed to have ergodic features.  [For the employed definition
or ergodicity see the discussion around (\ref{karamba2}, \ref{tam}); for
the precise definition of what do we mean by non-ergodicity see the
discussion around (\ref{ham_BO}).] Thus no microcanonical distribution
can be introduced for $\S$, unless the Born-Oppenheimer term is
neglected.  We show that together with the emergence of the
causal arrow, there appears a natural, microcanonical probability
distributions \footnote{$P(\F|\S)$ is the conditional probability for
the coordinates and momenta of $\F$, with the variables of $\S$ being
fixed.} $P(\S)$ and $P(\F|\S)$, where $P(\S)$ and $P(\F|\S)$ are simpler
(in the precise sense discussed below) than, respectively, $P(\F)$ and
$P(\S|\F)$. The above simplicity argument for the causal
reasoning thus gets validated in the present approach. 

In section \ref{fast_slow} we define the system to be studied. Sections
\ref{energy_F} and \ref{mixing} discuss, respectively, the dynamics of
the fast subsystem F and the convergence of its probability distribution
toward the microcanonic distribution. Dynamics of the slow subsystem S
is described in section \ref{dynamics_S}. In section \ref{trali} we
discuss in detail the (in)application of the basic statements of the
second law (thermodynamic arrow) to the dynamics of S. The joint
emergence of the thermodynamic arrow and the causal arrow is outlined in
section \ref{cau}. Section \ref{simplo} relates the obtained results to
the simplicity principle proposed recently in machine learning.  The
last section presents our conclusions and offers some speculations.

\section{Fast and slow subsystems.}
\label{fast_slow}

The overall Hamiltonian of $\S+\F$ reads
\BEA
\label{toto}
\CH(\Pi,Q,z)
=H_s(\Pi, Q)+H(z,Q),
\EEA
where $z=(q_1,...,q_N; p_1,...,p_N)$
are canonical coordinates and momenta of $\F$, and where
$Q=(Q_1,...,Q_M)$ and $\Pi=( \Pi_1,...,\Pi_M)$ are, respectively,
canonical coordinates and momenta of $\S$. The bare Hamiltonian of 
$\S$ is $H_s(\Pi, Q)$, while $H(z,Q)$ combines the bare Hamiltonian 
of $\F$ and the interaction Hamiltonian between $\S$ and $\F$.

Let $\tau_f$ be the characteristic time of $\F$ for the slow variable
$Q$ being fixed [for a more precise definition see after
(\ref{karamba2})]. We shall assume that both $Q$ and $\dot{Q}$ are slow
variables with respect to $\tau_f$. This assumptions is consistent with
the fact that the $\S-\F$ coupling involves only the coordinate $Q$ of
S: according to the Hamiltonian equation,
$\dot{Q}=\partial_\Pi[H_s(\Pi,Q)]$, generated by (\ref{toto}), $\dot{Q}$
does not depend explicitly on the fast variable $z$. 

Define $\nu_Q$ and $\nu_{\dot{Q}}$ as the characteristic times over
which $Q$ and $\dot{Q}$ change. Denote
\BEA
\label{times}
\tau_Q\equiv{\rm min}(\,\nu_Q, \, \nu_{\dot{Q}}\,).  
\EEA
Thus our basic
assumption on the separated time-scales (adiabatic limit) reads
\BEA
\label{adiabatic_limit}
\tau_f\ll \tau_Q.
\EEA

\section{Energy of the fast subsystem.}
\label{energy_F}

Our intention is to see how the energy $H(z,Q)$ of the fast
subsystem $\F$ changes in time.

Hamilton's equations of motion for the fast subsystem imply $\frac{\d
}{\d t}H(z_t,Q_t)=\dot{Q}_t\,\partial_Q H(z_t,Q_t)$.  Assuming the
adiabatic limit $\tau_{f}\ll\tau_Q$, and denoting $Q_t$ and $z_t$ for
the time-dependent coordinates, we have for the energy change on the
intermediate times $\tau_Q\gg \tau\gg \tau_f$:
\BEA
\label{hek}
\frac{\d}{\d \tau} E &&\equiv\frac{1}{\tau}[\,
H(z_{t+\tau},Q_{t+\tau})-H(z_t,Q_t)\,]\\
&&=\int_t^{t+\tau}\frac{\d s}{\tau}\,
\,\frac{\d H}{\d s}(z_s,Q_s)\\
\label{trek}
&&=\frac{\dot{Q}_t}{\tau}\int_t^{t+\tau}\d s\,
\partial_Q H (z_s,Q_t)+o(\frac{\tau}{\tau_Q}),
\EEA
where we took $\dot{Q}_t$ out of the integral, since $\dot{Q}_t$ (together
with $Q_t$) is assumed to be a slow variable. 

The last integral in (\ref{trek}) refers to the $Q={\rm const}$ dynamics
with $Q_t=Q$.  This dynamics has a constant energy $E=H(z,Q_t)$. 
Define for the microcanonic distribution
\begin{gather}
{\cal M}(z,E,Q)\equiv\frac{1}{\omega(E,Q)}\,\delta[E-H(z,Q)],\\
\omega(E,Q)
\equiv \int \d z\, \delta[E-H(z,Q)],
\label{in}
\end{gather}
where $\omega(E,Q)$ ensures the proper normalization: $\int\d z\, {\cal M}(z,E,Q)=1$.

Consider the following obvious relation:
\BEA
\label{karamba1}
\int \d z \,w(z){\cal M}(z,E)
=\frac{1}{\tau}\int_t^{t+\tau}\d s
\int \d z \,w(z){\cal M}(z,E),
\EEA
where $w(z)\equiv\partial_Q H(z,Q_t)$,
and where for simplicity we drop the explicit dependence on $Q=Q_t={\rm const}$.

In the RHS of (\ref{karamba1}) we change the integration variable as
$y={\cal T}_{t-s}\, z$, where ${\cal T}_{t}$ is the flow generated by
the Hamiltonian $H(z)=H(z,Q_t)$ between times $0$ and $t$.
Employing Liouville's theorem, $\d z=\d y$,
and energy conservation, ${\cal M}(z,E)={\cal M}(y,E)$, one gets
\BEA
(\ref{karamba1})=\int \d y\,{\cal M}(y,E)\,
\frac{1}{\tau}\int_t^{t+\tau}\d s \,w({\cal T}_{s-t}\,y).
\label{karamba2}
\EEA
If $w(z)$ is an {\it ergodic observable} of the $Q_t=$const dynamics,
then by definition of ergodicity there is such a characteristic time
$\tau_f$ such that for $\tau\gg \tau_f$ the time-average in
(\ref{karamba2}) depends on the initial condition $y$ only via its
energy $H(y,Q_t)$ \cite{vk,berdi}.  Since ${\cal M}(y,E)$ is
proportional to a $\delta$-function at $E=H(z,Q_t)$, the integration over
$y$ in (\ref{karamba2}) drops out, and we get that the time-average in
(\ref{karamba1}) is equal to the microcanonical average at the energy
$E$.  Applying this to the time-average in (\ref{trek}) we get
\BEA
\label{tam}
\frac{\d E}{\d\tau} = \frac{\d Q}{\d\tau} \int \d
z\,\partial_Q H(z,Q_t)\,{\cal M}(z,E_t,Q_t),
\EEA
where we noted again that $\dot{Q}$ is a slow variable.

We define the phase-space volume $\Omega$ enclosed by the energy shell $E$:
\BEA
\label{uganda}
\Omega(E,Q)
\equiv\int \d z\,\theta(E-H(z,Q)).
\EEA

Let us see how $\Omega(E,Q)$ changes in the slow time:
\BEA \label{der} 
\frac{\d}{\d\tau}
\Omega(E,Q)=\partial_E\Omega|_Q\,\, \frac{\d E}{\d\tau}+\partial_Q\Omega|_E \,\,\frac{\d Q}{\d\tau}.
\EEA
Using (\ref{tam}, \ref{uganda}) we get 
\BEA
\frac{\partial_Q\Omega|_E}{\partial_E\Omega|_Q}=-
\int \d z\,\partial_Q H(z,Q_t)\,{\cal M}(z,E_t,Q_t), 
\label{kabbala}
\EEA
and then from (\ref{tam}, \ref{der}, \ref{kabbala}):
\BEA
\frac{\d}{\d\tau}\Omega(E,Q)&=&\partial_E\Omega|_Q\left[
\frac{\d E}{\d\tau}+\frac{\d Q}{\d\tau}\, \frac{\partial_Q\Omega|_E}{\partial_E\Omega|_Q}
\right]\nonumber\\
&=&\partial_E\Omega|_Q\left[
\frac{\d E}{\d\tau}-\frac{\d E}{\d\tau}
\right]
=0.
\EEA

Thus, the phase-space volume $\Omega(E,Q)$ 
is an adiabatic invariant, i.e., it is conserved within the
slow dynamics. In particular, 
in the adiabatic limit the points of the fast phase-space located
initially at the energy shell $E_{\rm i}$ appear on the energy
shell $E_{\rm f}$, which is found from 
\BEA 
\label{tri}
\Omega(E_{\rm i}, Q_{\rm i})= \Omega(E_{\rm f}, Q_{\rm f}). 
\EEA
Since by definition (\ref{uganda}), $\Omega(E)$ is an increasing function of $E$,
for given $Q_{\rm i}, Q_{\rm f}$ and $E_{\rm i}$ the equation (\ref{tri})
has a unique solution
\BEA
\label{mri}
E_{\rm f}\equiv h(Q_{\rm f}|E_{\rm i}, Q_{\rm i}), 
\EEA

In the adiabatic limit the energy 
$h(Q_{\rm f}|E_{\rm i}, Q_{\rm i})$ of $\F$
does not depend on the precise phase-space location of the fast
trajectory on the energy shell $E_{\rm i}$. 

Note that the derivation of (\ref{der}) does not demand the 
full ergodicity|which means that {\it all} smooth observables of F are ergodic|only
certain observable is assumed to be ergodic \cite{vk}. The argument
expressed by (\ref{karamba1}, \ref{karamba2}) applies to calculating the
time-average of any ergodic observable $w(z)$ of $\F$ for a fixed $Q$. 

The adiabatic invariance of $\Omega$ for ergodic systems is well known
\cite{hertz,berdi,rugh} and motivated the microcanonic definition of
entropy as $\ln \Omega$ \cite{berdi,rugh}.  The
precision of the invariance is studied in \cite{ott}. We presented the 
above derivation for the completeness of this work and for highlighting
the two basic assumptions that are not properly articulated in literature: {\it i)}
ergodicity of an observable versus the full ergodicity, {\it ii)} and the
necessity for both $Q$ and $\dot{Q}$ being slow. 

\section{Conditional microcanonic distribution of the fast subsystem.}
\label{mixing}

For describing time-averages of ergodic observables of $\F$ (see
(\ref{karamba1}, \ref{karamba2}) and the discussion after (\ref{tam}))
we can employ the following time-dependent microcanonic conditional
probability:
\BEA
\label{gupta}
P_f[z|Q_{\rm i}, \Pi_{\rm i}]=
\frac{\delta[h(Q_\tau|E_{\rm i},Q_{\rm i})-H(z,Q_\tau)] 
}{\int \d z\, \delta[h(Q_\tau|E_{\rm i},Q_{\rm i})-H(z,Q_\tau)]}.
\EEA
Below we explain how to find $Q_\tau$ given the initial energy $E_{\rm i}$ 
of $\F$, the initial canonical coordinates $Q_{\rm i}$, $\Pi_{\rm i}$
of $\S$ and the time $\tau$. Note that $P_f[z|Q_{\rm i}, \Pi_{\rm i}]$ is 
time-dependent and varies with time on the slow time-scale $\tau\sim \tau_Q$.

There is another way of introducing the microcanonic distribution
(\ref{gupta}) which explicitly uses
the ensemble description \cite{zas,liebe}. If for a fixed $Q$ the system
$\F$ is {\it mixing}, then for any sufficiently smooth initial
probability distribution $p(z,0)$ of $\F$, the ensemble averages of
sufficiently smooth (i.e., sufficiently coarse-grained) observables
$A(z)$ of $\F$ converge in time to the averages taken over the
(\ref{gupta}) \cite{zas,liebe}: 
\BEA
\int \d z\, p(z,t) A(z)\to \int \d z\,P_f[z |Q,\Pi] A(z).
\EEA
The rate of this convergence defines the mixing time.
It is more natural (especially for chaotic systems) to define
observables via ensemble averages than via averages over time
\cite{zas}.  If not stated otherwise, 
from now on we assume that $\F$ is mixing, and thus
the mixing time coincides with $\tau_f(\ll\tau_Q)$ defined around
(\ref{karamba2}). 
For strongly (and homogeneously) chaotic systems the mixing time is
inversely proportional to the KS entropy \cite{zas,liebe}.

\section{Dynamics of the slow subsystem.}
\label{dynamics_S}

Let us average the equations of motion
$\dot{\Pi}=-\partial_Q[H_s(\Pi,Q)+H(Q,z)]$ and
$\dot{Q}=\partial_\Pi[H_s(\Pi,Q)]$ over the microcanonic distribution
(\ref{gupta}). We get that $\S$ is by itself a Hamiltonian system:
\BEA
\label{b1}
\dtau{\Pi}=-\partial_Q \CH_s,\quad
\dtau{Q}=\partial_\Pi  \CH_s ,
\EEA
with an effective Hamiltonian
\BEA
\label{ham_BO}
\CH_s(\Pi,Q|Q_{\rm i},E_{\rm i})
=H_s(\Pi,Q)+h(Q|Q_{\rm i},E_{\rm i}),
\EEA
which is the sum of $H_s(\Pi,Q)$ and the
Born-Oppenheimer term $h(Q|Q_{\rm i},E_{\rm i})$.  In
particular, $\CH_s(\Pi,Q|Q_{\rm i},E_{\rm i})$ determines the actual slow
trajectory $Q_\tau$, given its initial location $(\Pi_{\rm i}, Q_{\rm
i})$.  Substituting this back into (\ref{gupta}) we thus complete the
description of $\F$. 

The evolution generated by (\ref{b1}) conserves the energy $\CH_s$.  This is the
total energy of $\S+\F$ Note that the Born-Oppenheimer term $h(Q|Q_{\rm
i},E_{\rm i})$ depends on the initial coordinate $Q_{\rm i}$. This means
that the points in the phase-space $(\Pi,Q)$ that had initially equal
energy (but different initial coordinates $Q_{\rm i}$) will have
different energies at later times.  Thus $\S$ is not globally
Hamiltonian. 

While this fact seems to be of no special importance when we consider a
single slow trajectory, it matters much for developing statistical
physics for $\S$. 
Indeed, there is no global slicing of the
phase space into energy shells which makes the definition of the
microcanonic distributions impossible.

Thus $\S$ is non-ergodic: once ergodic systems are characterized by
loosing the memory on the initial phase-space location and remembering
only the initial energy (recall the argument around (\ref{karamba1},
\ref{karamba2})), in the considered situation the very form of the
energy depends on the initial phase-space location.

\subsection{Liuoville equation and Liuoville theorem.}

A consequence of the non-globally Hamiltonian dynamics 
is that the Liouville equation and the corresponding
theorem do not hold.  With the Hamilton equations (\ref{b1}) one can
relate a conditional probability
\BEA
&&\CP_{\rm con}(\Pi,Q,\tau|\Pi_{\rm i}, Q_{\rm i},0)\nonumber\\
&&=\delta(\Pi-\Pi(\Pi_{\rm i}, Q_{\rm i},\tau))\,\,
\delta(Q-Q(\Pi_{\rm i}, Q_{\rm i},\tau)),
\label{b2}
\EEA
where $\Pi(\Pi_{\rm i}, Q_{\rm i},\tau)$ and $Q(\Pi_{\rm i}, Q_{\rm i},\tau)$ are the solutions
of (\ref{b1}) with initial conditions $(\Pi_{\rm i}, Q_{\rm i})$.

As follows from (\ref{b1}, \ref{b2}),
$\CP_{\rm con}(\Pi,Q,\tau|\Pi_{\rm i}, Q_{\rm i},0)$ does satisfy to the Liouville equation
\BEA
\label{b3}
\partial_\tau \CP_{\rm con}=\partial_Q \CH_s\,\partial_\Pi \CP_{\rm con}
-\partial_\Pi \CH_s\,\partial_Q \CP_{\rm con}.
\EEA
Were $\CH_s$ not dependent on $Q_{\rm i}$, the direct integration of
(\ref{b3}) with the initial distribution $\CP(\Pi_{\rm i}, Q_{\rm i},0)$
would produce the Liouville equation for the unconditional probability
$\CP(\Pi, Q,\tau)$. But since $\CH_s(\Pi,Q|Q_{\rm i})$ does depend on
$Q_{\rm i}$, the integration with $\CP(\Pi_{\rm i}, Q_{\rm i},0)$ does
not lead to a differential equation for $\CP(\Pi, Q,t)$. 

Thus the Liouville equation and together with it the Liouville theorem
(conservation of the phase-space volume) do not hold. Below we shall
demonstrate this on an explicit example.

\subsection{An example.}
\label{example}

We assume that F and S without mutual coupling are 
two free particles, with masses $m$ and $M$, respectively. The 
S--F coupling
creates a harmonic potential for F:
\BEA
\label{q1}
H(p,q,Q)=\frac{p^2}{2m}+\frac{Q^2q^2}{2}.
\EEA
If we regard the slow variable $Q$ as a parameter, 
F is an ergodic system with the characteristic time
\BEA
\label{pino}
\tau_f=
\frac{2\pi\sqrt{m}}{Q}.
\EEA

Eq.~(\ref{der}) reduces to the conservation of action:
$E/|Q|={\rm const}$,
and thus the Born-Oppenheimer potential
$h(Q|E_{\rm i}, Q_{\rm i})$ reads from (\ref{tri})
\BEA
\label{q3}
h(Q|E_{\rm i}, Q_{\rm i})=E_{\rm i}\frac{|Q|}{|Q_{\rm i}|}.
\EEA

As the simplest example of the bare slow Hamiltonian we can take
free motion with a mass $M$:
\BEA
H_s=\frac{\Pi^2}{2M}.
\label{q4}
\EEA

Thus the dynamics of the slow subsystem S is described by the effective Hamiltonian:
$\CH_s=\frac{\Pi^2}{2M}+E_{\rm i
}\frac{|Q|}{|Q_{\rm i}|}$.
Assume that $Q>0$ and solve the Hamilton equations as:
\BEA
\label{terra}
\Pi(\tau)=\Pi_{\rm i}-\frac{E_{\rm i} \tau}{Q_{\rm i}},\quad
Q(\tau)=-\frac{E_{\rm i} \tau^2}{2MQ_{\rm i}}+\frac{\Pi_{\rm i}\tau}{M}+Q_{\rm i},
\EEA
where the initial time was taken $\tau=0$. The characteristic time $\nu_Q$ of $Q$
can be estimated from $Q(\nu_Q)-Q_{\rm i}\sim Q_{\rm i}$:
\BEA
\label{gusi}
\nu_Q={\rm min}\left[
\frac{M Q_{\rm i}}{\Pi_{\rm i}}, ~~\sqrt{\frac{2MQ^2_{\rm i}}{E_{\rm i}}}
\right].
\EEA
For the characteristic time $\nu_{\dot{Q}}$ of $\dot{Q}$ 
[estimated via $\dot{Q}(\nu_{\dot{Q}})-\dot{Q}_{\rm i}\sim \dot{Q}_{\rm i}$]
we get
\BEA
\label{krabo}
\nu_{\dot{Q}}
={Q_{\rm i}\Pi_{\rm i}}/{E_{\rm i}}.
\EEA
If $\Pi_{\rm i}\to 0$ we should take $\nu_{\dot{Q}}=
\sqrt{\frac{2MQ^2_{\rm i}}{E_{\rm i}}}$.

It is seen now that unless $Q(\tau)\simeq 0$, the adiabatic conditions
$\nu_{Q}\gg\tau_f$ and $\nu_{\dot{Q}}\gg\tau_f$ can be satisfied, e.g.,
for a sufficiently small $m$ and sufficiently large $M$. 

One now has from (\ref{terra}) for the Jacobian:
\BEA
\label{jacob} 
J(\tau)\equiv
\frac{\partial(\Pi(\tau),Q(\tau))}{\partial(\Pi_{\rm i},Q_{\rm i})}
=1-\frac{E_{\rm i}\tau^2}{2MQ^2_{\rm i}},
\EEA
which is not equal to $1$. Moreover, its absolute value can be both
larger or smaller than one, since it is not difficult to see that the
conditions $Q>0$ and $\frac{E_{\rm i}\tau^2}{2MQ^2_{\rm i}}>2$ can be
satisfied together. 

Perhaps the most visible consequence of the absence 
of the Liouville theorem is that the fine-grained
entropy
\BEA
\label{fine_entropy}
{\cal S}_{fg}[\tau]=
-\int \d \Pi\, \d Q\,
\CP(\Pi, Q,\tau)\ln \CP(\Pi, Q,\tau),
\EEA
of the slow subsystem is not anymore constant. Indeed, take a small
phase-space volume $v(0)$ and assume that $\CP(\Pi, Q,0)$ is constant
inside of this volume and equal to zero outside.  The fine-grained
entropy (\ref{fine_entropy}) is then ${\cal S}_{fg}[\tau]=\ln v(\tau)$,
where $v(\tau)$ is got from $v(0)$ under action of the flow generated by
Hamiltonian $\CH_s$. Thus, ${\cal S}_{fg}[\tau]-{\cal
S}_{fg}[0]=\ln\frac{v(\tau)}{v(0)}= \ln \left| J(\tau)\right|$ can both
increase and decrease in the course of time, as (\ref{jacob})
illustrates. 

When one can neglect the non-conservation of the phase-space volume?
Taking in (\ref{jacob}) $\tau\sim\tau[Q]$, and going in (\ref{gusi}) 
to the limit of a small $E_{\rm i}$ or a large $M$, we get
that the non-conservation of the phase-space volume can be neglected
|though $Q$ still changes significantly|if the fast energy $E_{\rm i}$ is much smaller
than the bare slow energy $\frac{\Pi_{\rm i}^2}{2M}$.

\section{Thermodynamic arrow for the slow subsystem. }
\label{2law2}\label{trali}

\subsection{Thomson formulation of the second law. }

How the second law applies to the effectively Hamiltonian, autonomous
slow subsystem $\S$?  The basic formulation of the second law is due to
Thomson: no work can be extracted from initially equilibrium system via
a cyclic change of an external field. This statement is derived as a
theorem both in classical and quantum mechanics \cite{Lindblad,Lenard}.
We already argued why this formulation is superior to the entropy
formulation: entropy is not directly observable and there is no general
consensus on its definition for a non-equilibrium state. In contrast,
work is directly observable, has a clear mechanical meaning, and its
general definition is universally accepted \cite{Balian,Lindblad}. Here
we focus on Thomson's formulation, while the entropic formulation is
studied below. 

Let us recall the statement of the Thomson formulation when no
interaction between $\S$ and $\F$ is present, i.e., 
the dynamics of S is generated by 
\BEA
\label{a1}
H_s(\G,\lambda_\tau), \qquad \G\equiv (Q,\Pi).
\EEA
The interaction of $\S$ with an external sources of work is described
by a time-dependent field $\lambda_\tau$ \cite{Balian,Lindblad}.

Let the initial phase-space points are sampled according to the Gibbs distribution:
\BEA
\CP_{G}(\G)=\frac{e^{-\beta H_s(\G)}}{Z},\quad
Z=\int \d \G\, e^{-\beta H_s(\G)},
\label{gi}
\EEA
where $\beta=1/T>0$ is the inverse temperature.
A cyclic change of the external field means:
\BEA
\lambda_0=\lambda_{\tau_c}=\lambda,
\label{mor}
\EEA
where $\tau_c$ is the cycle time. 

For the considered thermally isolated process 
the work is defined as the average energy difference
\footnote{\label{comrad}Work for a single trajectory $(\Pi_\tau,\, Q_\tau)$ is defined
as ${\cal W}=\int_0^\tau\d u\, \partial_{\lambda_u}
H_s(\Pi_u,Q_u,\lambda_u)
\frac{\d\lambda_u}{\d u}$.  
Employing the Hamilton equations of motion we get ${\cal W}=H_s(\Pi_\tau, Q_\tau, \lambda_\tau)
-H_s(\Pi_{\rm i},Q_{\rm i},\lambda_{\rm i})$, where $(\Pi_\tau, Q_\tau, \lambda_\tau)$
and $(\Pi_{\rm i},Q_{\rm i},\lambda_{\rm i})$ are the corresponding initial and final values.
Averaging this expression over the initial and final values, 
and recalling (\ref{mor}),
we get the expression of work 
as the average energy difference (\ref{2l}).}, 
and the statement of the Thomson formulation reads \cite{Lindblad,Lenard}:
\begin{gather}
W= \int \d \G\,
H_s(\G,\lambda)[
\widetilde{\CP}(\G,\tau_c)-\CP_G(\G)]\geq 0,
\label{2l}
\end{gather}
where $\widetilde{\CP}(\G,\tau_c)$ is the final (at $t=\tau_c$)
probability distribution
obtained from the initial Gibbsian probability distribution $P_G(\G)$
via the Liouville equation with the time-dependent Hamiltonian
(\ref{a1}).

The inequality in (\ref{2l}) is essentially based on three facts {\it
i)} initial and final Hamiltonians are the same due to (\ref{a1},
\ref{mor}); {\it ii)} the same Hamiltonian appears in the initial Gibbs
distribution; {\it iii)} the Liouville equation.

The easiest way to establish the validity of (\ref{2l})
is to employ the positivity of the relative entropy \cite{Lindblad}:
\BEA 
S[\widetilde{\CP}(\tau_c)|| {\CP_G} ]
\equiv
\int \d \G\, \widetilde{\CP}(\G,\tau_c)\ln\frac{\widetilde{\CP}(\G,\tau_c)}
{\CP_G(\G)}
\geq 0,
\label{relo}
\EEA
which holds for any probability distributions $\widetilde{\CP}(\G,\tau_c)$ and ${\CP_G(\G)}$.
Employing in (\ref{relo}) the conservation of the fine-grained entropy, 
$S_{fg}[\widetilde{\CP}(\tau_c)]=S_{fg}[ {\CP_G} ]$, due to the Liouville theorem, we get
\BEA
(\ref{relo})
=\int \d \G\, \left[\CP_G(\G)
-\widetilde{\CP}(\G,\tau_c)\right]\ln\CP_G(\G)\geq 0,
\label{pau}
\EEA
and then substituting (\ref{gi}) into $\ln\CP_G(\G)$
in (\ref{pau}) and recalling (\ref{mor})
we arrive at (\ref{2l}).

Let us now return to the slow subsystem $\S$ coupled to $\F$.  Now the
slow Hamiltonian is given by (\ref{ham_BO}) instead of (\ref{a1}).  At
the initial time both these Hamiltonians are equal modulo a factor
$E_{\rm i}$. We shall assume that the initial probability for $\Pi $ and
$Q$ is still given by (\ref{gi}), while initially the fast system always
starts with the same energy $E_{\rm i}$. For instance it is described by 
the microcanonic probability distribution (\ref{gupta}), and then the
overall initial distribution of S and F is the product of the above specified
marginal distributions for S and F.

Thus the overall distribution is not Gibbsian and the applicability of
the Thomson formulation to the overall system is not automatic.
The work is still given by the average energy difference (of the
slow subsystem, or, equivalently, of the total system) calculated
via the effective slow Hamiltonian (\ref{ham_BO}). 
This can be argued for exactly in the same way as in Footnote \ref{comrad}.
Instead of (\ref{2l}) we now get
\begin{gather}
\label{2ll}
W= \int \d \G\,
H_s(\G,\lambda)\left[
\CP(\G,\tau_c)-\CP_G(\G)\right]\\
+\int \d Q\, \d Q_{\rm i}
[h(Q|Q_{\rm i},E_{\rm i})-E_{\rm i}]
\CP(Q,\tau_c; Q_{\rm i},0),
\label{2lll}
\end{gather}
where $\CP(\G,\tau_c)$ is the phase-space probability distribution at $t=\tau_c$, while
$\CP(Q,\tau_c; Q_{\rm i},0)$ is the two-time probability distribution of the coordinate.
It is necessary to use the two-time distribution, since $h(Q|Q_{\rm i},E_{\rm i})$ explicitly
depends on both initial and final values of the coordinate.
Following to the steps outlined after (\ref{relo}) we get
\BEA
\label{misak1}
W&=&T\left(\,S[\CP(\tau_c)|| {\CP_G} ]+S_{fg}[ \CP(\tau_c) ]-S_{fg}[ {\CP_G} ]\,\right)
\\
&+&\int \d Q\, \d Q_{\rm i}
[h(Q|Q_{\rm i},E_{\rm i})-E_{\rm i}]
\CP(Q,\tau_c; Q_{\rm i},0),~~
\label{misak2}
\EEA
where the temperature $T$ comes from (\ref{gi}).  The first term
$TS[\CP(\tau_c)|| {\CP_G} ]$ in the RHS of (\ref{misak1}) is
non-negative. The fine-grained entropy difference $S_{fg}[ \CP(\tau_c)
]-S_{fg}[ {\CP_G} ]$ does not have definite sign, since the Liouville
equation does not hold. Moreover, the RHS of (\ref{misak1}), equal to
$T\int\d \Gamma\, [\CP_G(\G)-\CP(\G,\tau_c) ]\ln \CP_G(\G)$, does not
have a definite sign either.  Even if the
latter term is positive|e.g., because the fine-grained entropy increased
in time: $S_{fg}[ \CP(\tau_c) ]>S_{fg}[ {\CP_G} ]$|the term in
(\ref{misak2}) does not have any reason to be positive.  Apart of
special coincidences, there is no reason why the two ``dangerous'' terms
$S_{fg}[ \CP(\tau_c) ]-S_{fg}[ {\CP_G} ]$ and (\ref{misak2}) would
cancel each other. 

Thus the proof of Thomson's formulation can fail two times:
once because the Liouville equation does not hold, and second time
because a cyclic change (\ref{mor}) of the parameter $\lambda$ does not
yet imply a cyclic change of the Born-Oppenheimer term (this is the
origin of the term in (\ref{misak2})). 

The latter aspect can be studied separately.  Let S be a single
particle, and assume the following natural choice of the bare slow
Hamiltonian: $H_s(\Pi,Q)=\frac{\Pi^2}{2M}+V(Q)$, where the potential
$V(Q)$ has its deepest minimum at $Q_0$: $V(Q)> V(Q_0)$ for $Q\not=
Q_0$. In the initial Gibbs distribution of $\S$ take $T= 0$.
Then the initial distribution is reduced to a single initial condition
$\Pi_{\rm i}=0$ and $Q_{\rm i}=Q_0$. The interaction of S with external
sources of work is described by an additional potential
$u(Q,\lambda_\tau)$, which is equal to zero both initially and at the end
of the cycle; see (\ref{mor}). We assume that at intermediate times
$u(Q,\lambda_\tau)$ is such that $Q_0$ ceases to be a local minimum of
the overall potential, i.e., the particle located initially at $Q_0$
will move out of it and will change its energy. Now for the work one has
analogously to (\ref{misak1}, \ref{misak2}):
\BEA
\label{kobra2}
W&=&H_s(\Pi(\tau_c), Q(\tau_c))-H_s(0,Q_0)\\
&+&h(Q(\tau_c)|E_{\rm i},Q_0)-E_{\rm i},
\label{kobra1}
\EEA
where $\Pi(\tau_c)$ and  $Q(\tau_c)$ are the values of the canonical coordinates
at the end of the cyclic process. They are obtained from solving (\ref{b1},
\ref{ham_BO}). The term in (\ref{kobra1}) corresponds to that in (\ref{misak2}).

While $H_s(\Pi(\tau_c), Q(\tau_c))-H_s(0,Q_0)$ is non-negative by
construction, there is no general restriction on the sign of
$h(Q(\tau_c)|E_{\rm i},Q_0)-E_{\rm i}$.  
Noting the freedom in choosing $h(Q|E_{\rm i},Q_{\rm i})$, one can
make $h(Q(\tau_c)|E_{\rm i},Q_0)-E_{\rm i}$ so negative that the 
overall work is negative as well: $W<0$. 

\subsection{Entropic formulation of the second law.}

The invalidity of the entropic formulation is studied along similar
lines.  Assume that $\S$ consists of several subsystems:
$(\Pi;Q)=(\Pi_1,...,\Pi_M; Q_1,...,Q_M)$ (see Eq.~(\ref{toto})).
Coarse-grained entropy of $\S$ is defined as
\BEA
\label{ku}
{\cal S}_{cg}[\tau]= -\sum_{k=1}^M {\cal P}(\G_k,\tau)\ln {\cal P}(\G_k,\tau),
\EEA
where ${\cal P}(\G_k,\tau)$ is the corresponding one-subsystem
distribution function.  This is the sum of partial entropies for each
subsystem.  The difference ${\cal S}_{cg}[\tau]-{\cal S}_{fg}[\tau]$
between the coarse-grained entropy (\ref{ku}) and fine-grained entropy
(\ref{fine_entropy}) is non-negative (sub-additivity) and quantifies the
relevance of correlations in $\S$ \cite{Balian,Penrose}. 

For additionally motivating the definition (\ref{ku}), we can assume
that the subsystems of $\S$ were interacting for a finite time, and that
$\tau$ is larger than this interaction time. 

Note that the definition (\ref{ku}) is not the only possibility. There
are (infinitely) many ways of doing coarse-graining, and thus many ways
of defining non-equilibrium entropy \footnote{In particular, one can
focus on certain macroscopic observables and define their physical,
non-equilibrium entropy via maximization of information-theoretic
entropy \cite{Balian}.}.  The main advantage of (\ref{ku}) is that
allows to see the entropy increase due to correlations (which is the
main qualitative image behind the entropic formulation of the second
law) \cite{Balian,Penrose}. To this end assume that initially the
subsystems of $\S$ are independent
\BEA
{\cal P}(\G,0)=\prod_{k=1}^M{\cal P}(\G_k,0).  
\EEA
This assumption specifies initial conditions needed for the existence of
the thermodynamic arrow of time \cite{Balian,Penrose}. 
  
If $\S$ starts from such a non-equilibrium state, and if the
fine-grained entropy is constant in time due to the Liouville theorem,
then one employs sub-additivity to get that the coarse-grained entropy
is not decreasing in time
\begin{gather}
\label{o1}
S_{cg}(t)
\geq S_{fg}(t)=S_{fg}(0)=
S_{cg}(0).
\end{gather}
However, once the Liouville theorem is not satisfied, $S_{fg}$ can
decrease in time and then (\ref{o1}) does not hold in general.  There
are other schemes for deriving the entropic formulation of the second
law for different sets of initial states and for different definitions
of the non-equilibrium entropy \cite{Balian,Jaynes,Lindblad,Roeck}. All
these derivations essentially use the Liouville theorem, so that all of
them do not apply to the present situation. 

Note that there is a difference between inapplicability of the entropic
formulation as compared to that of the Thomson formulation. Eq.~(\ref{o1})
shows that if the fine-grained entropy increases in time, the entropic formulation
is satisfied. In contrast, the increasing fine-grained entropy does not yet
ensure the validity of the Thomson formulation, as we discussed after (\ref{misak2}).

\section{The causal arrow. }
\label{cau}

\subsection{Reciprocity versus negligibility of the Born-Oppenheimer term.}
\label{ganimed}

All the above anomalies with the second law are due to the fact that the
Born-Oppenheimer term $h(Q|Q_{\rm i},E_{\rm i})$ makes the dynamics of S
not globally Hamiltonian. There are two related options for recovering
this feature. First one can try to see whether the dependence of
$h(Q|Q_{\rm i},E_{\rm i})$ on $Q_{\rm i}$ can be neglected, $h(Q|Q_{\rm
i},E_{\rm i})\simeq h(Q|E_{\rm i})$, but $h(Q|E_{\rm i})$ still exerts a
sizable force on S.  Second, one can look for conditions where
$h(Q|Q_{\rm i},E_{\rm i})$ can be neglected as whole. We shall now show
that only the second option is consistent. 

Employing (\ref{tri}, \ref{mri}) as
\BEA 
\Omega(E_{\rm i}, Q_{\rm i})= \Omega( h(Q|E_{\rm i}, Q_{\rm i}), Q), 
\EEA
and using (\ref{in}) we get
\BEA
\label{zuk1}
\partial_{Q_{\rm i}} h(Q|E_{\rm i}, Q_{\rm i})=
\frac{\partial_{Q_{\rm i}}\Omega (E_{\rm i}, Q_{\rm i})
}{\omega(\,h(Q|Q_{\rm i},E_{\rm i}),\,Q)},\\
\label{zuk2}
\partial_{Q} h(Q|E_{\rm i}, Q_{\rm i})=-
\frac{\partial_{Q}\Omega (E, Q)|_{E=h(Q|E_{\rm i}, Q_{\rm i})
}}{\omega(\,h(Q|Q_{\rm i},E_{\rm i}),\,Q)}.
\EEA
These equations show that there is a certain reciprocity|to
be guessed already from (\ref{tri}, \ref{mri})|in the way $h(Q|E_{\rm i}, Q_{\rm i})$
depends on $Q$ and $Q_{\rm i}$. 

Let us demand that the Born-Oppenheimer term $h(Q|E_{\rm i}, Q_{\rm i})$
is independent from $Q_{\rm i}$. Since ${\omega(\,h(Q|Q_{\rm i},E_{\rm
i}),\,Q)}$ is finite, this demand amounts to $\partial_{Q_{\rm i}}\Omega
(E_{\rm i}, Q_{\rm i})\to 0$ for all $E_{\rm i}$ and $Q_{\rm i}$.  This
means requiring $\partial_{Q}\Omega (E, Q)|_{E=h(Q|E_{\rm i}, Q_{\rm
i})}\to 0$.  Due to (\ref{zuk2}), this implies that $h(Q|Q_{\rm
i},E_{\rm i})$ reduces to a constant $h(Q|Q_{\rm i},E_{\rm i})=E_{\rm
i}$, and|in addition|the energy of F does not change in time.  We are
thus led to assuming that there is no relevant interaction between S and
F, a trivial option which is definitely out of our interest. 

We are thus left with the second option:
for the time-scales relevant for the dynamics of $\S$ the Born-Oppenheimer 
Hamiltonian $h(Q|Q_{\rm i},E_{\rm i})$ in
(\ref{ham_BO}) is negligible compared to the bare slow Hamiltonian
$H_s(\Pi,Q)$. For this it is necessary to have:
\BEA \label{delta} H_s(\Pi,Q) \gg h(Q|Q_{\rm
i},E_{\rm i}). 
\label{hippo}
\EEA
In the absence of the Born-Oppenheimer term, the dynamics driven by $H_s$ is globally
Hamiltonian, the Liouville theorem holds, and the second law is
applicable to $\S$; see the previous sections. 

Using (\ref{tri}, \ref{mri}) and (\ref{in}) one calculates:
\BEA
\partial_{E_{\rm i}} h(Q|E_{\rm i}, Q_{\rm i})=
\frac{\omega(E_{\rm i},Q_{\rm i})}{\omega(h(Q|Q_{\rm i},E_{\rm i}),Q)}>0.
\label{frog}
\EEA
This means that the Born-Oppenheimer term decreases with $E_{\rm i}$. Since
the RHS of (\ref{frog})
is normally $\sim {\cal O}(1)$, for satisfaction of (\ref{hippo})
we have to require 
\BEA
H_s(\Pi,Q)\gg E_{\rm i}. 
\label{bela}
\EEA
We already saw this condition at the end of section \ref{example} for a
particular example. This example also shows that there may be
situations, where for sufficiently long times of the slow motion the
Born-Oppeheimer force cannot be neglected, even though it is numerically
small; see (\ref{jacob}) in this context. In addition, there can be time
limitations related to the validity of the time-scale separation, and
thus to the definition of the Born-Oppenheimer force; see the discussion
after (\ref{krabo}) in this context. Thus, at the moment we cannot give
a fairly general estimate for the times on which the conditions
(\ref{hippo}) and (\ref{bela}) will be sufficient for neglecting 
the Born-Oppenheimer term.

\subsection{The causal arrow.}

Eq.~(\ref{hippo}) also means that the interaction between $\S$ and $\F$
gets the causal arrow: $\S$ (cause) influences on $\F$ (effect), while
$\F$ does not influence on $\S$. 

Thus we see that for the present system, the thermodynamic arrow and the
causal arrow emerge simultaneously.  Recall in this context the
operational definitions of the causal arrow discussed in section
\ref{causal_intro}. 

\section{Microcanonical ensemble and simplicity principle. }
\label{simplo}

After neglecting the Born-Oppenheimer term $h(Q|Q_{\rm i},E_{\rm
i})$ we recover a globally Hamiltonian behavior for the dynamics of $\S$.
In particular, the time-average of the ergodic observables of $\S$ can
be described by probability distribution:
\BEA
\label{kappa}
P_s(\G)=\frac{\delta(U_s-H_s(\G))}{\int \d \G\,\delta(U_s-H_s(\G))},
\EEA
where $U_s$ is the slow energy. Since $\S$ does not get back-reaction
from $\F$, the energy $U_s$ is a constant determined by the initial
conditions for the dynamics of $\S$. 

Recall that the very existence of (\ref{kappa}) is related to
neglecting the back-action of $\F$ on
$\S$. For the same reason the probability distribution (\ref{kappa}) is
unconditional.  The appearance of (\ref{kappa}) can be argued following
to the lines of section \ref{mixing}. In this context we should assume
that $\S$ with the Hamiltonian $H_s(\Pi,Q)$ is mixing and define the
mixing time $\tau_s$ of $\S$. 

The distributions (\ref{gupta}) and (\ref{kappa}) can be combined into a
non-equilibrium microcanonic ensemble for describing the statistics of
the overall system $\S+\F$ on the times larger than $\tau_s$, but
smaller than the mixing time $\tau_{s+f}$ of the overall system:
\BEA
P(\G,z)=
P_s(\G) P_f(z|\G).
\label{combi}
\EEA
It is understood that $Q_\tau$ needed in (\ref{gupta}) for
defining $P_f(z|\G)$ is obtained (for given initial
$\G=(Q,\Pi)$) by solving the equations of motion
(\ref{b1}) for $\S$ without the Born-Oppenheimer term. 

Note that $P(\G,z)$ in (\ref{combi}) can be obtained via sequential
maximization of the conditional entropy $-\int\d z\, P(z|\G)\ln P(z|\G)$
of $\F$ for fixed slow variables, and then maximization of the
unconditional entropy $-\int\d \G\, P(\G)\ln P(\G)$ of $\S$ for fixed
slow energy $U_s$.  In this context it is not difficult to accept the
idea that the microcanonic distribution is the simplest (least
informative) one for a fixed value of energy. 

On the other hand, the probability distributions $P(\G|z)$ and
$P(z)$|obtained from (\ref{combi}) via the Bayes formula|are not simple.
They are not microcanonic, and in general they cannot be even obtained
in a closed form. 

Recalling that under condition (\ref{delta}) we identified $\S$ and $\F$
as the cause and effect, respectively, we get that the probability
distributions $P(\S)$ and $P(\F|\S)$ are simpler than $P(\F)$ and
$P(\S|\F)$. As proposed in Ref.~\cite{Jan}, in causal reasoning one should
tend to prefer the causal hypothesis $\C \rightarrow \E$ ($\C$ is the
cause, and $\E$ is its effect) if the factorization of $P(\C,\E)$ into
$P(\C) P(\E|\C)$ leads to significantly simpler terms $P(\C)$ and
$P(\E|\C)$ than the factorization into $P(\E) P(\C|\E)$.  Thus this
simplicity argument for the causal reasoning is validated in
the present approach. 

The causal arrow persists in the global microcanonic
equilibrium which|if the overall system $\S+\F$ is mixing with a time 
$\tau_{s+f}$|is established for $t\gg \tau_{s+f}$:
\BEA
P_{eq}(\G,z)=\frac{\delta({\cal E}-H_s(\G)-H(Q,z)    )
}{\int \d \G\, \d z\,\delta({\cal E}-H_s(\G)-H(Q,z)    )},
\label{globo}
\EEA
where ${\cal E}$ is the total energy. Eq.~(\ref{globo}) is a 
stationary distribution. The no-back-action condition
(\ref{hippo}) is now substituted by its equilibrium analog 
\BEA
H_s(\Pi,Q)\gg H(Q,z). 
\EEA
However, once the slow Hamiltonian is much larger than the fast Hamiltonian, we
expect that the partial probability $P_{eq}(\Pi,Q)$ will be close to
$P_s(\Pi,Q)$ in (\ref{kappa}). Indeed, once $H(Q,z)$ is small, the
overall energy ${\cal E}$ in (\ref{globo}) should be nearly canceled by
the bare slow Hamiltonian $H_s(\Pi,Q)$, so that $P_{eq}(\Pi,Q)$ is
proportional to a smeared delta-function concentrated at ${\cal
E}=H_s(\Pi,Q)$. For calculating observables (at small
$H(Q,z)$) this is the same as $P_{eq}(\Pi,Q)\propto \delta({\cal
E}-H_s(\Pi,Q))$. 

As for the conditional probability $P_{eq}(z|\Pi,Q)=P_{eq}(z|\G)$, 
it can always be written as
\BEA
P_{eq}(z|\G)=\frac{\delta({\cal E}-H_s(\G)-H(Q,z)    )
}{\int \d z\,\delta({\cal E}-H_s(\G)-H(Q,z)    )}.
\label{globo1}
\EEA
Here ${\cal E}-H_s(\G)$ is, of course, not the Born-Oppenheimer energy
$h(Q_\tau|E_{\rm i},Q)$ that shows up in the non-equilibrium
distribution (\ref{gupta}). Still ${\cal E}-H_s(\G)$ can be seen as an
equilibrium analog of $h(Q_\tau|E_{\rm i},Q)$. 

\section{Summary.}

We studied a Hamiltonian system that consists of a slow subsystems S and
a fast subsystem F; see section \ref{fast_slow}.  The separation into
slow versus fast is one of the basic ways of defining autonomous systems
in natural sciences \cite{Haken}. In particular, the effective dynamics
of slow subsystems is studied in a great variety of different fields:
atomic and molecular physics, semi-classic physics (including semi-classic
gravity), physical chemistry, synergetics, economics, {\it etc}. 

Our main purpose was in relating two seemingly different issues: {\it
i)} the causal arrow|or unidirectional influence|where S influences F,
but does not get back-action; {\it ii)} the thermodynamic arrow of time
(second law) for the system. Since the applicability of the second law
to F is well known \cite{Lenard,mwp}, we focused on the second law as
applied to the autonomous, energy conserving, Hamiltonian dynamics of S.
The presence of F is reflected in the dynamics of S via an additional
Born-Oppenheimer term in the Hamiltonian of S. This term emerged during
the tracing out of F, and it depends on the initial coordinate of S; see
section \ref{dynamics_S}. Thus, different initial coordinates of S have
different Hamiltonians: the dynamics of S is not globally Hamiltonian.
The cause of this is that due to the time-scale separation
the dynamics of F does have an adiabatic invariant
(effective conservation law); see section \ref{energy_F}. 

The specific features of the Born-Oppenheimer term make the basic
formulations of the second law inapplicable to the dynamics of S. These
statements of the second law
are {\it i)} the Thomson formulation, which states that no work can be
extracted by means of a cyclic Hamiltonian process (driven by an
external source of work), if the initial conditions of S are thermal and
{\it ii)} entropic formulation, which claims that the coarse-grained
entropy of S does not decrease, provided that S starts from a
low-entropy state. There are two mechanisms for this inapplicability.
First, the Liouville theorem (i.e., conservation of the fine-grained
entropy) does not hold for a non-globally Hamiltonian dynamics: the
fine-grained entropy can both increase or decrease in the course of
time.  The second mechanism is efficient for the Thomson formulation
only and has to do with the behavior of the Born-Oppenheimer term under
a cyclic Hamiltonian driving; see section \ref{trali} for details. 

As we argued in section \ref{ganimed}, the Born-Oppenheimer term has a
certain reciprocacy feature. Its basic implication for our purposes is
that the only way to recover a globally Hamiltonian dynamics for S is to
neglect the Born-Oppenheimer term as compared to the bare Hamiltonian of
S. By this we neglect the influence of F on S, but, importantly, the
influence of S on F is not neglected and can be sizable. Once
the Born-Oppenheimer term can be neglected, the basic formulations
of the second law naturally apply to S. Thus we see that the emergence of the
thermodynamic arrow (second law) for S is closely related to the causal arrow:
S acts on F, but does not get back-action. 

Finally, in section \ref{simplo}, we studied our results in the context
of a causal inference principle proposed recently in machine learning
\cite{Jan}.  This principle plausibly infers the causal-effect relation
between two stochastic variables, and it intends to cover especially
those situations, where more standard causal inference procedures do not
apply.  If we assume that S and F are mixing systems, under the causal
arrow condition they are described by the microcanonic probability
distribution $P(\S)$ and the conditional microcanonic distribution $P(\S
|\F)$. Now the factorization of the joint probability $P({\rm
cause=S},\, {\rm effect=F})$ into $P({\rm cause} )\,P({\rm effect}|{\rm
cause} )$ leads here to simpler expressions than the factorization into
$P({\rm effect} )\,P({\rm cause}|{\rm effect} )$. This is the core of
the inference principle proposed in \cite{Jan}, and we conclude that
this principle is validated in the present approach. 

\acknowledgements The work was supported by Volkswagenstiftung grant
``Quantum Thermodynamics: Energy and information flow at nanoscale''.

\end{document}